\documentclass[useAMS,usenatbib]{mn2e}

\usepackage{graphicx}

\title[Imaging search for the companion to $\epsilon$ Ind A]{Imaging search for the unseen companion to $\epsilon$ Ind A -- Improving the detection limits with 4 $\mu$m observations \thanks{Based on observations collected at the European Southern Observatory, Chile (ESO No.\ 082.D-0251 and 081.C-0430).}}
\author[M. Janson et al.]
{
M. Janson$^{1}$\thanks{Reinhardt fellow. E-mail: janson@astro.utoronto.ca}, 
D. Apai$^{2}$, 
M. Zechmeister$^{3}$, 
W. Brandner$^{3}$, 
M. K\"urster$^{3}$, 
\newauthor
M. Kasper$^{4}$, 
S. Reffert$^{5}$, 
M. Endl$^{6}$, 
D. Lafreni\`ere$^{1}$, 
K. Gei{\ss}ler$^{3}$, 
\newauthor
S. Hippler$^{3}$, 
Th. Henning$^{3}$
\\
\\
$^{1}$Department of Astronomy, University of Toronto, 50 St George St, Toronto, M5S 3H4, Canada\\
$^{2}$Space Telescope Science Institute, 3700 San Martin Drive, Baltimore, MD 21218, USA\\
$^{3}$Max Planck Institute for Astronomy, K\"onigstuhl 17, Heidelberg, D-69117, Germany\\
$^{4}$ESO, Karl-Schwarzschild-Strasse 2, Garching bei M\"unchen, D-85748, Germany\\
$^{5}$Landessternwarte, K\"onigstuhl 12, Heidelberg, D-69117, Germany\\
$^{6}$McDonald Observatory, 1 University Station, Austin, TX 78712, USA}
\begin{document}

\date{N/A}

\pagerange{\pageref{firstpage}--\pageref{lastpage}} \pubyear{2009}

\maketitle

\label{firstpage}

\begin{abstract}
$\epsilon$ Ind A is one of the nearest sun-like stars, located only 3.6 pc away. It is known to host a binary brown dwarf companion, $\epsilon$ Ind Ba/Bb, at a large projected separation of 6.7$'$, but radial velocity measurements imply that an additional, yet unseen component is present in the system, much closer to $\epsilon$ Ind A. Previous direct imaging has excluded the presence of any stellar or high-mass brown dwarf companion at small separations, indicating that the unseen companion may be a low-mass brown dwarf or high-mass planet. We present the results of a deep high-contrast imaging search for the companion, using active angular differential imaging (aADI) at 4 $\mu$m, a particularly powerful technique for planet searches around nearby and relatively old stars. We also develop an additional PSF reference subtraction scheme based on locally optimized combination of images (LOCI) to further enhance the detection limits. No companion is seen in the images, although we are sensitive to significantly lower masses than previously achieved. Combining the imaging data with the known radial velocity trend, we constrain the properties of the companion to within approximately 5-20 $M_{\rm jup}$, 10-20 AU, and $i > 20^{\rm o}$, unless it is an exotic stellar remnant. The results also imply that the system is probably older than the frequently assumed age of $\sim$1 Gyr.
\end{abstract}


\section{Introduction}

\label{sec_intro}

Direct imaging of exoplanets is a field of research in rapid development. The past year has seen a number of interesting planet candidates imaged directly around stars. Most notable is arguably HR 8799, showing three planetary companions (see Marois et al. 2008) so far. These planets have been shown to exhibit Keplerian motion around the star, and have estimated masses in the range of 7-10 $M_{\rm jup}$ from theoretical models (Baraffe et a. 2003). The system is known to also host a debris disk (e.g. Moor et al. 2006; Rhee et al. 2007). Along with the fact that there are three almost equal-mass companions, all orbiting the star in what appears to be a co-planar manner, this implies that the objects most likely formed in a circumstellar disk, by a process distinct from the star or brown dwarf formation process. This can be contrasted with the case of the 2M1207 system (Chauvin et al. 2005), where the low primary-to-secondary mass ratio is more reminiscent of a brown dwarf binary system than a star-planet system. Two other intriguing planet candidates in systems with debris disks were reported around the same time as HR 8799 b/c/d: Fomalhaut b (Kalas et al. 2008) and Beta Pic b (Lagrange et al. 2009), though additional follow-up observations would be desirable to provide more information on these systems. 

$\epsilon$ Ind A is a K4V-type southern sky star located at a distance of 3.6 pc, with a very high proper motion of 4.7$''$ yr$^{-1}$ (Perryman et al. 1997), see Table \ref{indtab0}. Comoving on the sky at the same rate, at a separation of 6.7$'$, is $\epsilon$ Ind B, which was first detected by Scholz et al. (2003), and shortly thereafter resolved into the brown dwarf binary $\epsilon$ Ind Ba/Bb (McCaughrean et al. 2004) as it is known today. Being the most nearby binary brown dwarf, and with a physical separation small enough to determine its dynamical mass within a reasonable timeframe, $\epsilon$ Ind Ba/Bb will be a benchmark object for the physical understanding of brown dwarfs. However, in addition to the Ba/Bb components, the $\epsilon$ Ind system may provide an additional possibility to study an even lower-mass and cooler object. $\epsilon$ Ind A displays a linear radial velocity trend (Endl et al. 2002; Zechmeister et al., in prep.) which, unless due to some exotic stellar remnant, is indicative of a giant planet or very low-mass brown dwarf companion. If this object could be directly imaged, it would constitute yet another important benchmark object, given its probable low mass, low temperature, and the possibility to estimate both its luminosity as well as its dynamical mass within a reasonable timeframe. It would also be the closest planet or very low-mass brown dwarf companion directly detected outside of our Solar System.

\begin{table}
   \centering
\caption[]{Summary of $\epsilon$ Ind A properties.}
         \label{indtab0}
\begin{tabular}{ll}
            \noalign{\smallskip}
            \hline
            \noalign{\smallskip}
                 Property & Value \\
            \noalign{\smallskip}
            \hline
            \noalign{\smallskip}
Right Ascension  & 22 03 21.66 \\
Declination   & -56 47 09.5 \\
Spectral Type  & K4.5V \\
Distance & $3.626 \pm 0.009$ pc \\
Proper Motion  & 4705 mas yr$^{-1}$  \\
Mass & 0.7 $M_{\rm sun}$  \\
Age & 1-5 Gyr \\
            \noalign{\smallskip}
            \hline
\end{tabular}
\end{table}

Here, we will present our deep imaging campaign of $\epsilon$ Ind, using narrow-band 4$\mu$m imaging with two different high-contrast techniques. One technique is pure active angular differential imaging (aADI) as already implemented and demonstrated as a powerful technique for detecting close companions to bright nearby stars (Janson et al. 2008). In aADI (also known as roll subtraction), images are taken at two different instrument rotator angles, and one is subtracted from the other, thereby removing the bulk of the stellar PSF including static instrumental speckles. The other technique is a combination of PSF reference subtraction and aADI (PSFR+aADI), using the LOCI (locally optimized combination of images) algorithm developed by Lafreniere et al. (2007). We also compare the performance of aADI at 4$\mu$m with the same technique in the L'-band. The concept of using aADI with L' was proven by Kasper et al. (2007; the usefulness of L' for high-contrast imaging purposes was also independently demonstrated by Hinz et al. 2006). The concept of using NB4.05 to enhance the physical contrast was introduced in Janson et al. (2008).

The outline is as follows: In Sect. \ref{sec_obs}, we summarize the observational parameters and ambient conditions of the observing runs. The two paths of data reduction employed are described in Sect. \ref{sec_datared}. This is followed by a presentation of the results and the associated analysis in Sect. \ref{sec_results}, including a comparison between filters (Sect. \ref{sec_filters}), a comparison between techniques (Sect. \ref{sec_methods}), a discussion of what we can learn from the dynamical input in Sect. \ref{sec_dynamics}, and the final detection limits and their interpretation in Sect. \ref{sec_detlimits}. Finally, we conclude in Sect. \ref{sec_conclusions}.

\section{Observations}

\label{sec_obs}

The data presented here are based on two different sets of VLT/NACO observations of $\epsilon$ Ind. One set of observations consisted of deep imaging with aADI in the NB4.05 filter, executed in service mode and split into two equal observing blocks (henceforth observations A1 and A2), on 31 Oct 2008 and 2 Nov 2008. The other set of observations were taken in visitor mode on 3 Jul 2008, as part of a larger survey searching for planets around a volume-limited sample of nearby stars (Apai et al., in prep.). Those observations were less deep, but consisted of aADI imaging in both the NB4.05 and broad-band L' filters (henceforth observations B1 and B2, respectively). For observations A1 and A2, the same strategy was used as in previous observations (see Janson et al. 2008): Jittering was applied to enable a good subtraction of the thermal background, and the aADI was performed at two different instrument rotator angles, using a differential angle of 33 degrees. For B1 and B2, a four-point large-throw dithering scheme was applied for the background subtraction purposes, and the differential angle used for aADI was 20 degrees. All observations were taken with the L27 objective, providing a pixel scale of 27 mas/pixel, and a field of view of 28$''$ by 28$"$. The weather conditions and observational parameters of each run are listed in Table \ref{indtab1}.

\begin{table}
   \centering
\caption[]{Observational conditions for the four runs.}
         \label{indtab1}
\begin{tabular}{lllll}
            \noalign{\smallskip}
            \hline
            \noalign{\smallskip}
                 ~ & A1   & A2   & B1 & B2\\
            \noalign{\smallskip}
            \hline
            \noalign{\smallskip}
Date (2008)         & 31 Oct & 2 Nov & 3 Jul & 3 Jul \\
Filter          & NB4.05 & NB4.05 & NB4.05 & L'  \\
Seeing$^a$    & 0.9$''$    & 0.7$''$    & 1.1$''$ & 1.0$''$  \\
Strehl$^b$    & 83\%      & 79\%      & 85\% & 83\% \\
Humidity      & 8\%         & 13\%        & 3\% & 3\%  \\
Coh. time$^a$ & 2.6 ms      & 3.4 ms      & 2.3 ms & 2.4 ms \\
Frames             & 19          & 19$^c$          & 30 & 33 \\
(per angle)        &      ~      &     ~       &       ~  & ~  \\
DIT                & 1.0 s       & 1.0 s       & 0.2 s & 0.2 s \\
NDIT               & 61          & 61          & 150  & 150  \\
Tot. time          & 1159 s      & 1159 s$^c$      & 900 s & 990 s  \\
(per angle)        &      ~      &     ~       &       ~  & ~  \\

            \noalign{\smallskip}
            \hline
\end{tabular}
\begin{list}{}{}
\item[$^{\mathrm{a}}$] Values given by the atmospheric seeing monitor at a wavelength of 500 nm.
\item[$^{\mathrm{b}}$] Strehl ratio given by the AO system, rescaled to the observing wavelengths.
\item[$^{\mathrm{c}}$] Two frames were de-selected for 0$^{\rm o}$, hence the effective time is 1037 s for that case.
\end{list}
\end{table}

\section{Data reduction}

\label{sec_datared}

The data reduction for A1 and A2 was done differently with respect to the two different techniques applied, hence separate descriptions are provided below. The reduction of B1 and B2 was only done with aADI.

\subsection{Reduction for the purpose of aADI}

\label{sec_aadired}

Since observations A1 and A2 were taken in the same way as the observations of Janson et al. (2008), largely the same data reduction could be applied as in the case for aADI purposes: The most basic reduction steps (e.g. flat fielding, bad pixel removal, background subtraction) were provided by the ESO automatic pipeline. The subsequent steps were performed with our dedicated IDL pipeline: The images were shifted using bilinear interpolation to a common center determined through cross-correlation, and to an absolute center using center of gravity. Low-frequency filtering was applied by subtracting a smoothed counterpart of each image produced by convolution with a Gaussian kernel with 0.5$''$ FWHM (the FWHM of the stellar PSF was about 120 mas). All the images corresponding to different rotation angles were subtracted from each other, and the results from the two nights were coadded. As an alternative analysis, all images were also de-rotated back to a common angle and coadded. This procedure yields a single co-added signature of any companion, which provides a useful alternative way to look at the data in the background-limited regime, with respect to the aADI-subtracted data which instead produces two independent signatures of half the amplitude each.

For observations B1 and B2, the data reduction was performed with the IDL routines as described above, with the exception that the centering was determined based on a sub-frame of 200x200 pixels around the star instead of the whole frame. This was done in order to avoid influence from residual features from the sky subtraction and the different dithering scheme.

The residuals as function of separation from the star were calculated from the standard deviation of all pixels in an annulus corresponding to each separation step. The physical brightness contrast was derived from the 3$\sigma$ residuals by division of the peak value of the primary. Since the primary was saturated in the science images (the saturation radius was about 5-6 pixels), this had to be calibrated. For runs A1 and A2, this was done by introducing a neutral density filter into the optical path during acquisition, thus getting non-saturated images of the primary, allowing to determine a renormalized peak value. For run B1, the same non-saturated images as for A1 and A2 were used, which could be done since the Strehl ratio was stable and almost equal between the epochs. For B2, an image was taken of a fainter photometric standard star during the night.

\subsection{Reduction for the purpose of PSFR+aADI}

\label{sec_psfrred}

In applications involving LOCI, it is preferable to maximize the number of PSF representations of a target or reference star, hence for this case, reduction was done on individual frames, with the combination of frames only performed at the very end. The PSF star used was $\epsilon$ Eri, which is practically ideal for the purpose given the similar spectral type, brightness, and the fact that observations exist taken under almost identical circumstances. All individual target and reference star frames were manually subjected to flat fielding, dark subtraction and bad pixel correction using calibration frames provided by ESO. Low-frequency filtering was applied as described above. A master sky frame was then produced by taking the median of the frames, where the stellar image is randomly placed in each frame, thereby removing the star altogether. The individual frames revealed ring-like structures in the background that could be reproduced in the master sky frame. By subtracting the master sky frame from each individual frame, the pattern could be removed. The pattern was found to be constant during the extent of an observation, but variable between observations (e.g., the frames corresponding to $\epsilon$ Ind and $\epsilon$ Eri were different from each other), and is probably related to instrumental dust emitting at 4$\mu$m. The full background subtraction obtained in this way was found to be equally good as that delivered by the ESO 'jitter' routine. It is interesting to note, that given the fact that the pattern is constant during an observation, it should be possible to calibrate it out of a generic observation by making a master sky observation directly before or after the target observation. Hence, for any observation dedicated to the detection of point sources, it should be possible to achieve the same degree of background subtraction with and without jittering. This is an important realization with respect to high-contrast imaging at these wavelengths using techniques such as passive ADI (pADI) or coronagraphy, where it is desirable to maintain the stellar primary at a fixed position on the detector, and simultaneously achieve the best possible background subtraction. In summary, there appears to be no conflict between these two requirements, as long as the master sky calibration step is performed during observations.

PSFR and aADI were performed separately, in sequence. For PSFR, every target and reference image was de-rotated such that the spider patterns were aligned. For each target frame, an optimized PSF reference frame was then produced from the full set of reference frames using the LOCI (Lafreniere et al. 2007) algorithm and subtracted from the target frame. The optimization was performed in 10 regions, five for the image range contaminated by the four spiders, covering different radial sections of the PSF, and five for the image range not contaminated by spiders, also covering different radial sections. The spider optimization areas were rectangular with a fixed width of 25 pixels, inner radii of 10, 20, 40, 70, and 120 pixels, and outer radii of 60, 70, 90, 150, and 200 pixels. The remaining areas were annuli excluding the spider regions, between inner radii of 20, 30 40, 50, and 60 pixels and outer radii of 50, 60, 70, 80, and 100 pixels. The subtractions were performed sequentially outwards with the subtraction zone defined from the inner radius of the optimization zone and outwards. Following this procedure, each of the target frames were re-rotated to their true parallactic angle. The aADI step was then performed through a second LOCI PSF construction, using all 33$^{\rm o}$ frames as PSF library for each 0$^{\rm o}$ frame, and vice versa. For this case, the optimization regions were simply five annuli between inner radii of 10, 20, 30, 50, and 70 pixels and outer radii of 40, 50, 60, 70 and 100 pixels. The optimization regions were chosen to provide a good balance between the two main criteria of LOCI: to maximize the efficiency of stellar PSF structure subtraction, and minimize subtraction of actual companions. The latter was tested by generating a series of runs where false companions had been introduced in the target frame -- in total 3600 companions distributed between 10 and 100 pixels separation from the center of the star, and over all azimuthal angles. The partial subtractions in each case were used to construct a radial profile of conserved companion flux fractions. As expected, a significant flux loss occurs at 10 pixels, but decreases rapidly outwards. At 100 pixels, the fraction of restored companion flux approaches unity, as indeed expected, given that the LOCI optimization is not applied beyond 100 pixels for the vast majority of the image space. 

Finally, all frames corresponding to each rotator angle were combined using 3$\sigma$-clipping. The radial profile of residuals was created in the same way as for aADI, but with the additional step that it was normalized by the radial profile of conserved companion flux fraction to provide an accurate measure of the actual achieved contrast.

\section{Results and discussion}

\label{sec_results}

\begin{figure*}%[htb]
   \centering
   \includegraphics[width=15cm]{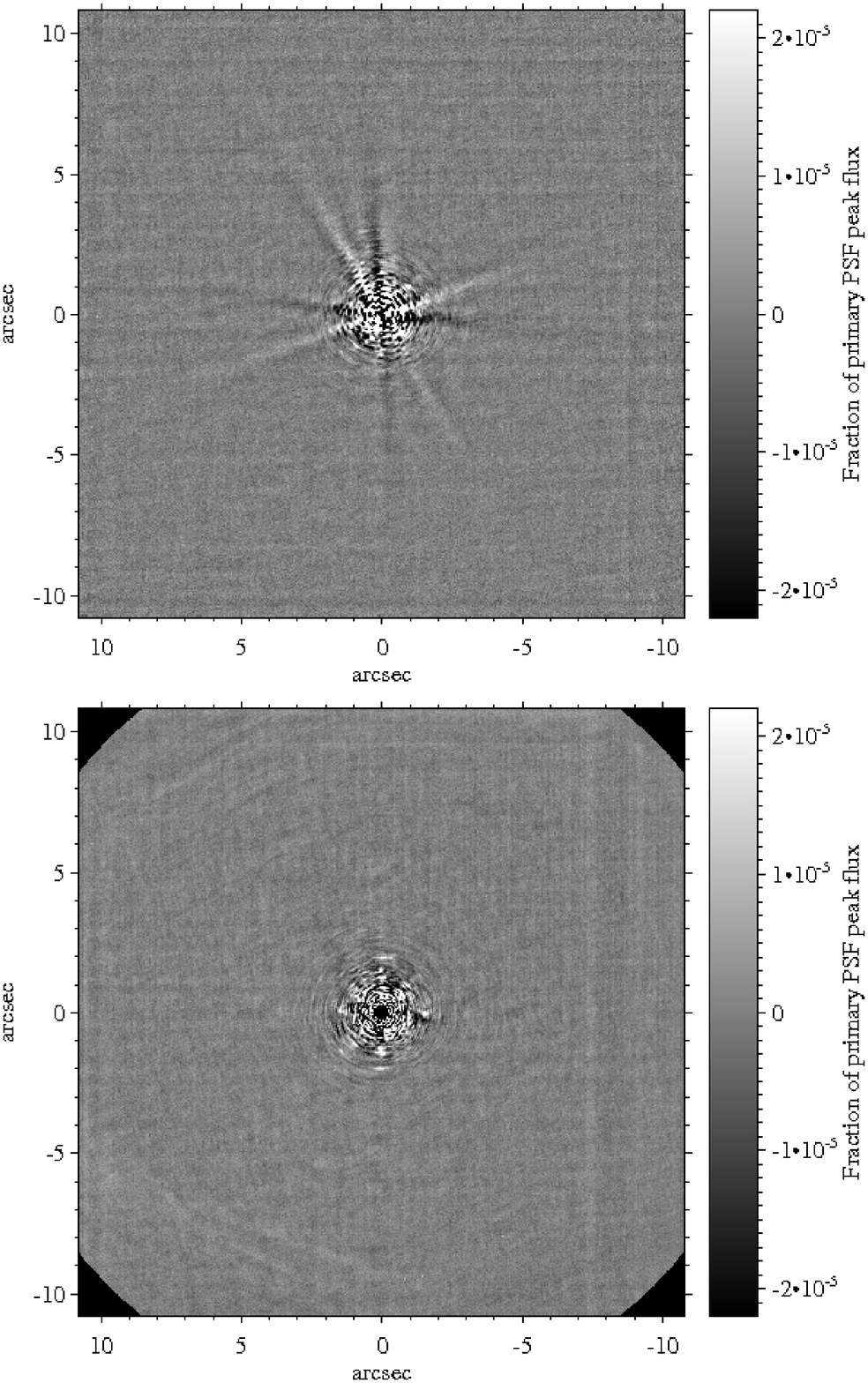}
\caption{Final output images for aADI (top) and PSFR+aADI (bottom) to the same flux scale, for the images A1+A2. The most obvious difference between the methods is in the treatment of the spiders.}
\label{ei_imgs_mrg}
    \end{figure*}

The output images from runs A1+A2 from each of the two reduction paths are shown in Fig. \ref{ei_imgs_mrg}. No companion candidates were detected in the images. In the following, we discuss the implications of this result, and compare the methods used.

\subsection{Comparison between filters}

\label{sec_filters}

Although the B1 and B2 images are less deep than A1+A2, the fact that they were obtained for the same target at about the same time, and with an almost identical observational setup, makes them ideal for comparing L' and NB4.05 imaging for planet detection purposes around bright stars. A comparison was already made in Janson et al. (2008) between L' aADI, NB4.05 aADI, and SDI+aADI (from Janson et al. 2007). While a fully relevant comparison could be made between SDI+aADI and NB4.05 aADI, where NB4.05 aADI was found to perform better under all circumstances, the comparison with L' was preliminary, since no comparable data was available. Instead, the comparison between L' and NB4.05 was based entirely on physical contrast given by the theoretical models, and the instrumental contrast was assumed to be the same. While this is relevant for a large part of the parameter space, there will in reality be differences in instrumental contrast due to differences in Strehl ratio, PSF diffraction, and thermal background between the filters. Using the B1 and B2 observations, we can now provide a comparison that takes all these issues into account.

\begin{figure}%[htb]
   \centering
   \includegraphics[width=8.0cm]{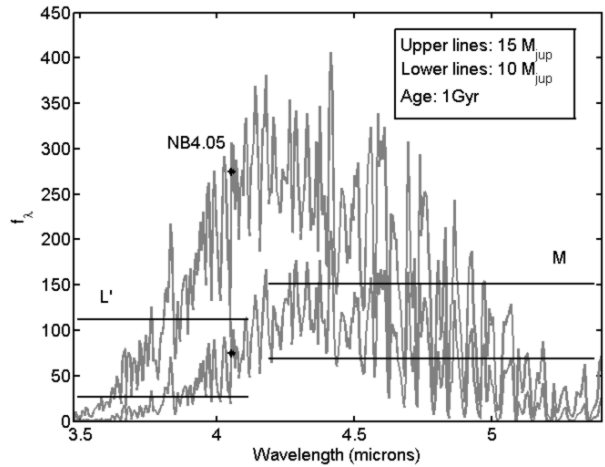}
\caption{Example of two model spectra from Burrows et al. (2003), and the corresponding flux densities in filters L', NB4.05, and M. Upper lines: A 15$M_{\rm jup}$ object. Lower lines: A 10$M_{\rm jup}$ object. The age is 1 Gyr in both cases.}
\label{xmpl_spec_sm}
    \end{figure}

The comparison was done by translating the brightness contrasts into mass detection limits using the spectral and photometric evolutionary models of Baraffe et al. (2003) and Burrows et al. (2003) for various ages. The method is described in detail in Janson et al. (2008). Note that the comparison is done for almost identical observing time, and with virtually no difference in overheads, i.e. the telescope time investment is also the same in both cases. As expected, the instrumental contrast is almost identical in the contrast-linited range, confirming the assumptions of the previous analysis, and thus the difference in the inner range is almost entirely set by the expected flux distribution of the companion. We show an example that demonstrates the favourable spectral range of NB4.05 in Fig. \ref{xmpl_spec_sm}, for 10$M_{\rm jup}$ and 15$M_{\rm jup}$ objects, at an age of 1 Gyr. The flux density is higher in NB4.05 than in both L'-band and M-band. For cooler objects, the bulk of the flux moves redward, hence M-band becomes better in terms of flux density, but the thermal background is also much worse in M-band. The improvement of NB4.05 over L'  increases further for cooler objects.

\begin{figure}%[htb]
   \centering
   \includegraphics[width=8.0cm]{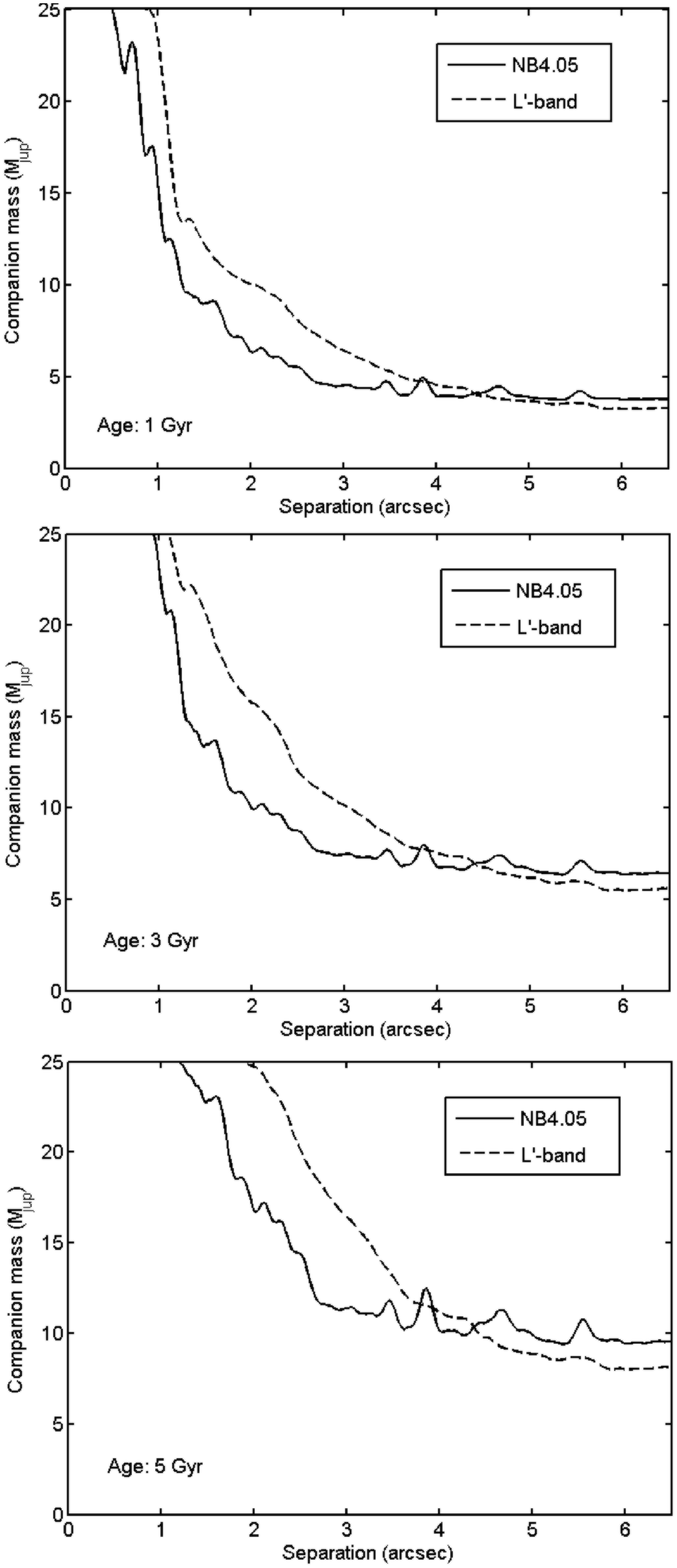}
\caption{Comparison between L' (dashed line) and NB4.05 (solid line) imaging for high-contrast purposes, for 1 Gyr (top panel), 3 Gyr, (middle panel), and 5 Gyr (bottom panel). As expected, NB4.05 provides a better performance than L' in the inner image range, and the opposite is true in the outer range. The comparison is based on sets B1 and B2, note that the A1+A2 detection limits are better.}
\label{brg_vs_lband}
    \end{figure}

We show the results of the detection limit comparison for 1 Gyr, 3 Gyr, and 5 Gyr in Fig. \ref{brg_vs_lband}. It is seen that for all these ages, NB4.05 performs better in the contrast-limited inner part, and L' performs better in the outer background-limited part, as expected. The crossover point for $\epsilon$ Ind A in our dataset is at about 4$''$. The position of the crossover point will vary as a function of stellar brightness and integration time. The brighter the star and the longer the integration time, the larger the parameter range where NB4.05 will be favourable,  and vice versa. We conclude that NB4.05 is likely to be an excellent choice for very deep planet search imaging close to bright stars, although it should be noted that this depends on the validity of the theoretical models. A first test of the models could be provided by the HR 8799 system.

\subsection{Comparison of aADI and PSFR+aADI}

\label{sec_methods}

\begin{figure}%[htb]
   \centering
   \includegraphics[width=8.0cm]{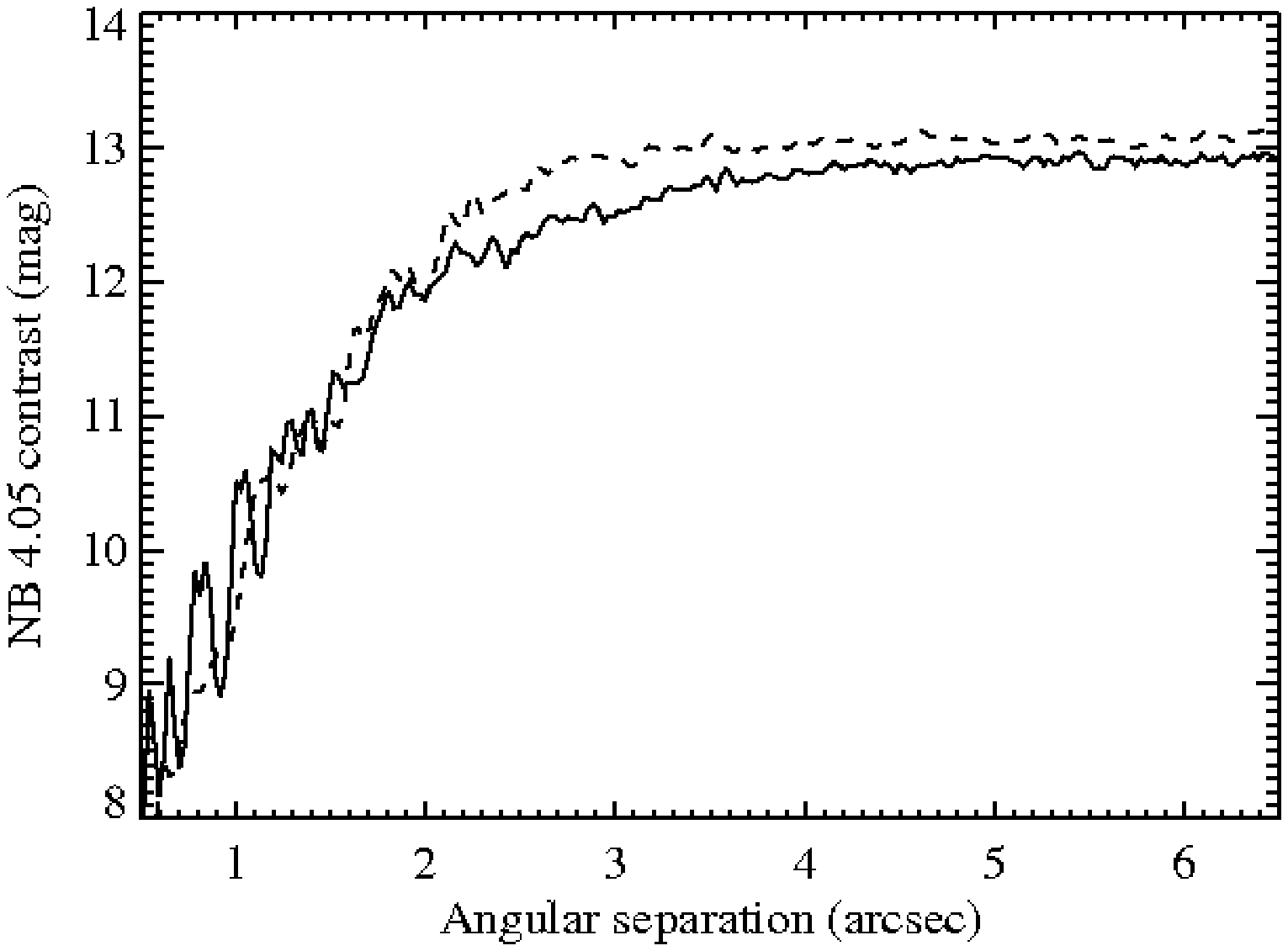}
\caption{Comparison of contrast for aADI (solid line) and PSRF+aADI (dashed line). The performance is generally very similar.}
\label{comp_aadi_loci}
    \end{figure}

As can be seen in the images (Fig. \ref{ei_imgs_mrg}), the main difference between aADI and PSFR+aADI is that spiders are more efficiently removed in the latter case. However, the impact of this is largely cosmetic, as a comparable amount of flux is lost from the companion in the spider regions. This can be clearly seen in a comparison of the respective contrast curves for the two methods (see Fig. \ref{comp_aadi_loci}), which have been normalized with respect to flux losses. PSFR+aADI slightly improves the performance at large separations, but provides no improvement at all for small separations. It should be noted that the results are based on a single PSF reference star (though with multiple representations) -- it would be preferable to use multiple reference stars, and doing so might substantially improve the performance. In any case, we do not reach as promising results as those achieved with PSFR using LOCI on space-based HST data (see Lafreniere et al. 2009), where a significant improvement over aADI is readily seen. As we have demonstrated, 4 $\mu$m imaging provides a very high Strehl ratio, so if this was the limiting PSF stability factor at this level of contrast, we should have expected an improvement in the inner image range. Hence, the results imply that other PSF effects become dominant once the Strehl ratio is high enough, such as low-order aberrations arising in the telescope, and differences in PSF representation resulting from dithering. This in turn implies that a stable telescope configuration is the best way forward for improving the contrast in 4 $\mu$m imaging even further. There is an obvious and well-tested technique for achieving this, called passive ADI, in which the pupil is stabilized during observations, while the field is allowed to rotate (see Marois et al. 2006). Indeed, the LOCI algorithm was originally designed for this purpose (Lafreniere et al. 2007). In fact, we have a passive ADI sequence at 4 $\mu$m showing exquisite performance at small separations, but those data are taken with a different telescope and of a different target, so a rigorous comparison can not be made. The passive ADI data will be part of a separate publication.

\subsection{Input from dynamics}

\label{sec_dynamics}

There exist extensive radial velocity measurements of $\epsilon$ Ind A, as well as some limited astrometric information, which can be used to constrain the properties of any sufficiently massive companion, as discussed in the following.

\subsubsection{The radial velocity linear trend}

\label{sec_rvtrend}

The linear radial velocity trend of $\epsilon$ Ind A was first reported by Endl et al. (2002). The original dataset covered an observational baseline of about 5.2 years, taken with the ESO CES instrument in the period 1992-1998. Since then, HARPS data have been taken from 2003 to 2008 (Zechmeister et al., in prep.). The linear trend of the HARPS data is consistent with that of the aforementioned CES data, with a slope of 4.4 m s$^{-1}$ yr$^{-1}$. Hence, we adopt this slope over the 16 year total baseline. One contributor to the linear trend is secular acceleration. This is the apparent acceleration that an observer measures over time in projected motion (in this case along the line of site) of an object with constant velocity, due to the actual motion in 3D space. Using all the measured spatial coordinates and velocity components of $\epsilon$ Ind, the secular acceleration can be calculated to 1.8 m s$^{-1}$ yr$^{-1}$. This is quite large, due to the fast motion of $\epsilon$ Ind in the plane of the sky, but still leaves a 2.6 m s$^{-1}$ yr$^{-1}$ trend that must be due to actual acceleration.

Since $\epsilon$ Ind Ba/Bb is known to be physically bound to $\epsilon$ Ind A, it needs to be tested whether it could be responsible for the observed trend. We do this with the following order-of-magnitude estimate: The projected separation between A and Ba/Bb is about 1500 AU, hence for masses of 0.7$M_{\rm sun}$, 0.047$M_{\rm sun}$, and 0.028$M_{\rm sun}$ respectively (see McCaughrean et al. 2004), the orbital period of the A/B system is at least 66000 years for a circular Keplerian orbit. Such an orbit would lead to a radial velocity semi-amplitude for $\epsilon$ Ind A of 62 m s$^{-1}$, which in turn gives an average peak-to-peak acceleration of $4*10^{-3}$ m s$^{-1}$ yr$^{-1}$. Thus, the gravitational influence of $\epsilon$ Ind Ba/Bb is several orders of magnitude too small to make any significant contribution to the observed trend.

With $\epsilon$ Ind Ba/Bb out of the picture, we are left with closer, as of yet unseen companions. A previous H- and K-band imaging campaign (Gei{\ss}ler et al. 2007) has excluded the presence of stellar and massive brown dwarf companions, down to 53 $M_{\rm jup}$ outside of a projected separation of 1.5 AU and 21 $M_{\rm jup}$ outside of 4.7 AU. This also excludes white dwarfs, since at ages up to several Gyrs, they are much brighter in H-and K-band than a 50$M_{\rm jup}$ object (see e.g. Holberg \& Bergeron 2006 and Baraffe et al. 2003). Stellar objects outside of the field of view can be excluded, as they would be detectable with wide-field or all-sky surveys such as 2MASS (Skrutskie et al. 2006). While more exotic forms of stellar remnants (e.g. neutron stars) can perhaps not be categorically excluded, for the remainder of this paper we will assume that the observed acceleration is due to a low-mass brown dwarf or giant planet. The combined constraints from the imaging and the radial velocity trend are given in Sect. \ref{sec_detlimits}.

\subsubsection{Astrometry}

\label{sec_astrometry}

As will be seen in the following, the companion is expected to have a mass in the range of $\sim$5-20$M_{\rm jup}$, and an orbital semi-major axis in the range of $\sim$10-20 AU. At the distance of the  $\epsilon$ Ind system, this corresponds to a strong astrometric amplitude signature imposed on the primary of about 15-60 mas. However, with an orbital period of a few decades, it would not be possible to detect orbital motion with, e.g., \textit{Hipparcos} data alone. On the other hand, one might expect a systematic difference between the proper motion as measured by \textit{Hipparcos} versus that measured in long-term ground based monitoring, such as from the Fifth Fundamental Catalog (FK5). This type of signature is referred to as $\Delta \mu$ binarity, see Wielen et al. (2001). For $\epsilon$ Ind, an approximate conversion between the FK5 and HIP systems implies that there is a difference between the \textit{Hipparcos} and FK5 proper motions of $\Delta \mu _{\alpha} = -0.23 \pm 1.68$ mas yr$^{-1}$, and $\Delta \mu _{\delta} = -2.5 \pm 0.98$ mas yr$^{-1}$. This corresponds in total to a significance level of $F=2.54$, where the $F$ value is roughly to the same level of confidence as the equivalent $\sigma$-number for Gaussian statistics (Wielen et al. 2001). In other words, there is an indication of a companion in the data, but not at a very high level of significance. We can make an order-of-magnitude estimation of whether these numbers are consistent with the RV companion by assuming that the orbital motion is completely averaged out in the FK5 data, that the orbit is circular, and that a sufficiently small fraction of the orbit is covered by \textit{Hipparcos} such that local curvature in the motion during that period is negligible. The limiting cases quoted above then yield astrometric motions of $\pi*15$ mas in 32 years and $\pi*60$ mas in 89 years respectively, i.e. 1.5 mas yr$^{-1}$ and 2.0 mas yr$^{-1}$, both of which are consistent with the given $\Delta \mu$ within the errors. Hence, the astrometry is indeed consistent with the RV trend, though we reiterate that the significance is rather limited for the astrometry.

\subsection{Detection limits}

\label{sec_detlimits}

\begin{figure}%[htb]
   \centering
   \includegraphics[width=8.0cm]{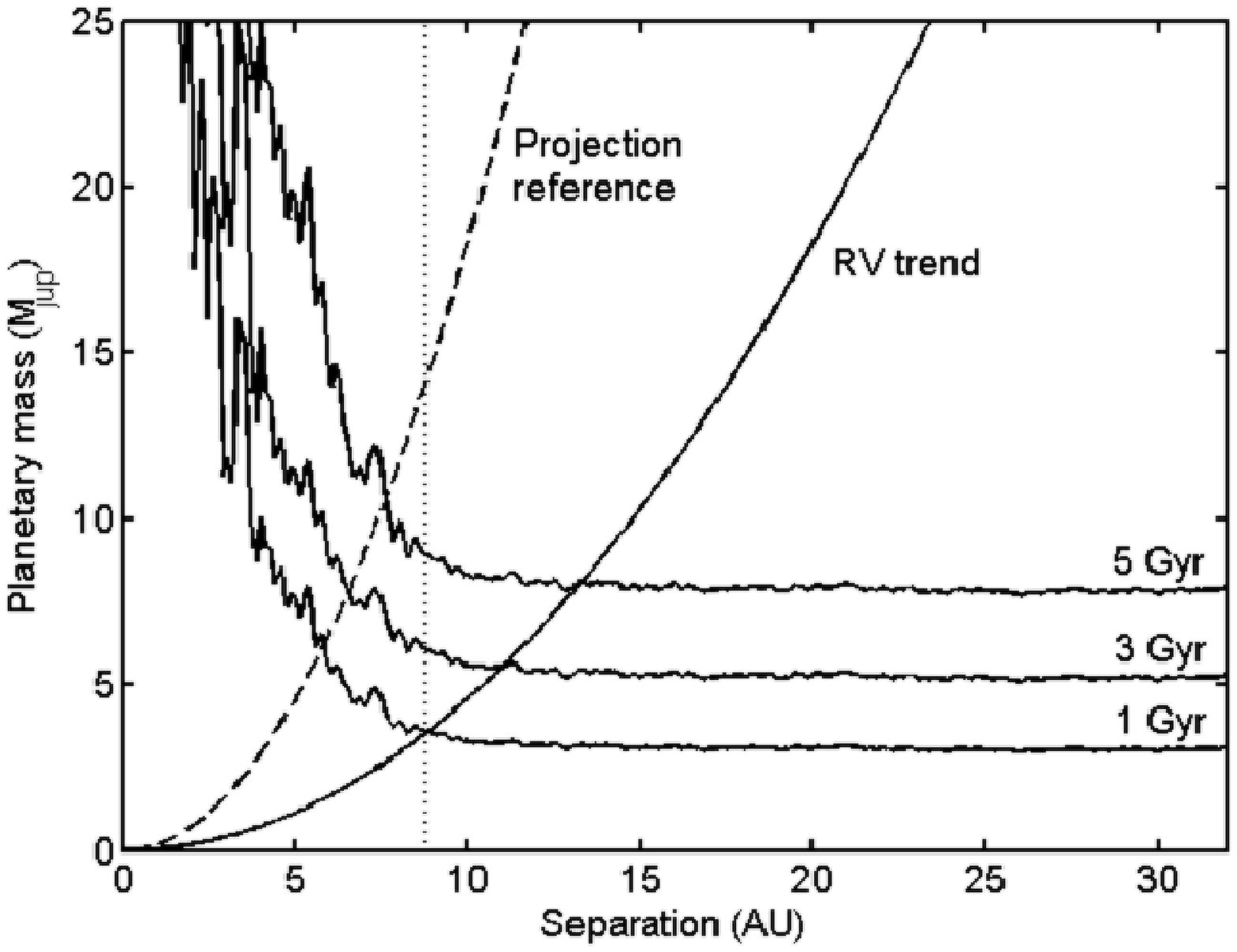}
\caption{Detection limits at 1, 3, and 5 Gyr for $\epsilon$ Ind A. The solid line that increases outwards is the mass as function of semi-major axis corresponding to the 2.6 m s$^{-1}$ yr$^{-1}$ slope of the observed RV trend, and the dashed line is the reference for minimum projected separation at a typical inclination of 60$^{\rm o}$, both under the assumption of a circular orbit. The dotted vertical line is the minimum semi-major axis from the RV baseline at $q=2$.}
\label{mlim_rv}
    \end{figure}

Since A1+A2 are the deepest images, they provide the strongest detection limits, and therefore we concentrate on them in this section. The instrumental contrast for A1+A2 is determined at each separation as the maximum performance out of the aADI and PSFR+aADI contrasts at that separation. The corresponding mass limits for A1+A2, calculated in the same way as for B1 and B2 in section \ref{sec_filters}, are shown in Fig. \ref{mlim_rv} for ages of 1, 3, and 5 Gyr. Several age determinations exist pointing to an age in the range of 1 Gyr for $\epsilon$ Ind (e.g. Lachaume et al. 1999; Barnes 2007). However, preliminary analysis of the astrometric masses of $\epsilon$ Ind Ba/Bb (Cardoso et al. 2008) implies that the components are probably under-luminous with respect to model predictions (Baraffe et al. 2003) at 1 Gyr, such that the $\epsilon$ Ind system has to be significantly older, perhaps up to 5 Gyr, if the models are accurate (which may not be the case, see e.g. Dupuy et al. 2009). On the other hand, such an old age would be incompatible with the observed spectra of $\epsilon$ Ind Ba and Bb according to the analysis of Kasper et al. (2009). It is with these uncertainties in mind that we consider the full range of 1 to 5 Gyr in our analysis.

Also plotted is the mass as function of semi-major axis derived from the slope of the linear trend, under the assumption that the inclination is $60^{\rm o}$ (the mean inclination of randomly oriented orbits). The minimum possible semi-major axis is set by the minimum possible period, which in turn is some multiple $q$ of the observational baseline. The exact value of $q$ depends on the amount of curvature present in the trend, the determination of which would be an over-interpretation of the data at hand. As discussed in Janson et al. (2008), $q=1$ would be the most conservative limit possible to set, but it is unrealistic, since it would require a discrete change in velocity state. Here, we set $q=2$, which is still conservative, and more realistic.

The mass limits and RV trend shown in Fig. \ref{mlim_rv} provide a good illustration of the detectability of the dynamical companion under the assumption of a circular orbit. However, given the large eccentricity spread of the exoplanet population outside of 0.1 AU, it is necessary to perform more detailed simulations in order to constrain the possible physical and orbital parameters of the companion. The method for doing so is described in detail in Janson et al. (2008), and we follow it here for $q=2$. In brief, based on the empirical distribution of eccentricities for known exoplanets outside of 0.1 AU, we simulate all possible orbits and orbital phases and test whether they are consistent with the observed linear trend. The fraction of such orbits as function of semi-major axis is named $\phi$. Out of these allowed orbits, we test what fraction would lead to a detectable companion. This fraction as a function of semi-major axis is termed $\chi$. One addition has been made to this procedure with respect to what was presented in Janson et al. (2008): In the case of $\epsilon$ Eri, the plane of the disk, the rotational plane of the star, and the orbital plane of the planet candidate $\epsilon$ Eri b all gave a consistent orbital inclination of about 30$^{\rm o}$, hence this number was fixed in the simulations. In the case of $\epsilon$ Ind, we have no prior information of the inclination, hence it is treated as a free parameter in the simulations. This is done by performing the simulations over several different inclination angles and averaging the results. The input inclination angles are set to correspond to the actual probability of a given inclination occurring (i.e., accurately taking into account that the inclination is more likely to be edge-on than face-on). 

   \begin{figure}%[htb]
   \centering
   \includegraphics[width=8.0cm]{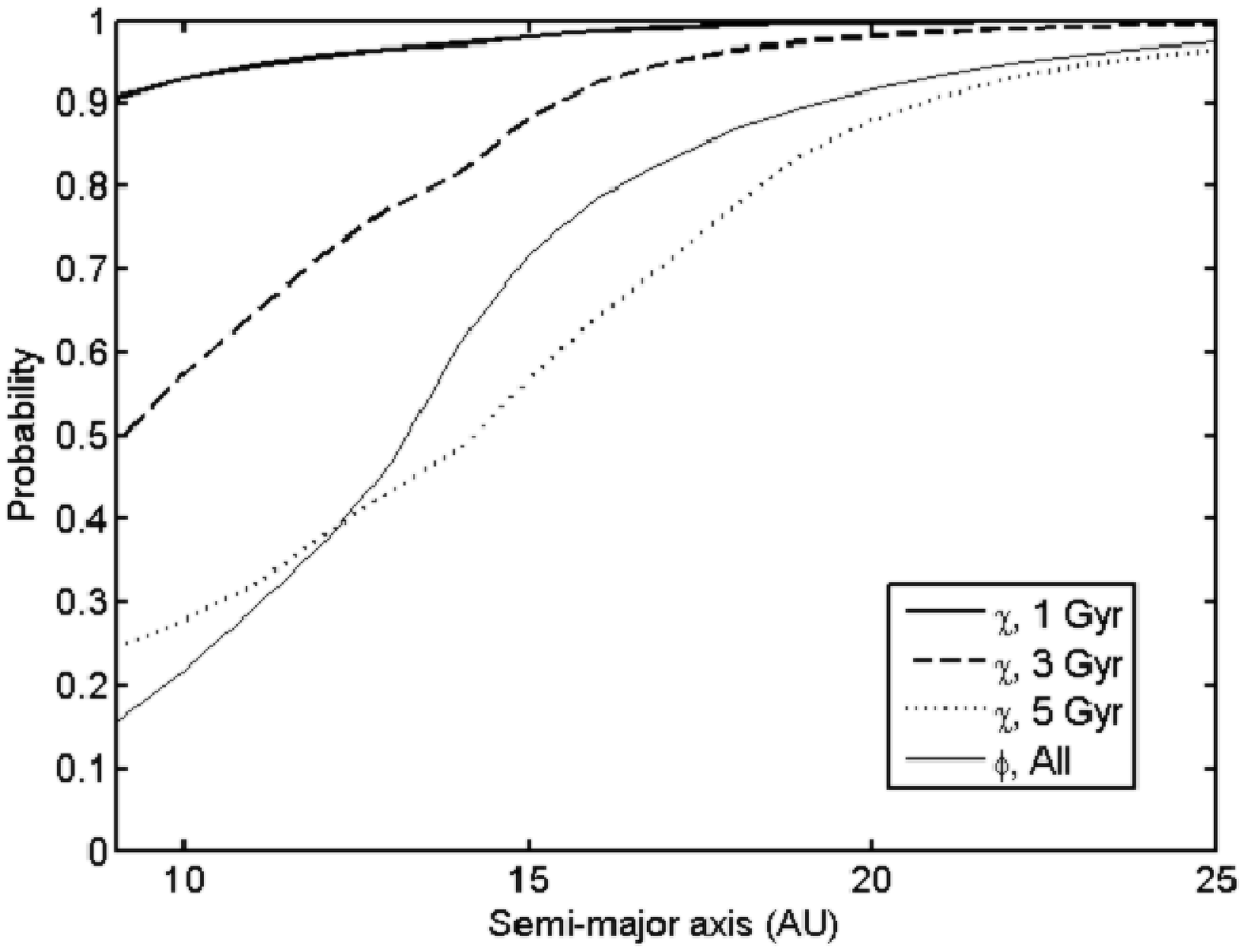}
\caption{Detection probability in our images as function of semi-major axis for 1, 3, and 5 Gyr. Also plotted is $\phi$, the fraction of orbits at a given semi-major axis that are consistent with the linear RV trend, denoted with the subscript "All" to signify that it is independent of age, in contrast to $\chi$.}
\label{probfig_q2}
    \end{figure}

The results of the simulation are shown in Fig. \ref{probfig_q2}. It can be seen that if the age is 1 Gyr, the probability of detecting the companion is always about 90\% or higher for any semi-major axis, hence since no companion is detected, it is quite unlikely that the system is that young. On the other hand, if the age is 3 Gyr, or even 5 Gyr as discussed above, there is still a substantial parameter range in which the companion could hide. Given these results, in approximate numbers we can constrain the planet or brown dwarf mass to about 5-20 $M_{\rm jup}$ and its semi-major axis to about 10-20 AU. Also, the inclination must be larger than at least 20$^{\rm o}$, otherwise the projection effects could never bring the companion close enough to the star to hide it, and the actual mass would be sufficiently larger than the projected mass to make it brighter than the background at any reasonable age.

\section{Conclusions}

\label{sec_conclusions}

We have attempted to image the indirectly discovered companion to $\epsilon$ Ind A, using imaging in the 4$\mu$m filter as well as the L'-band. As expected, 4$\mu$m imaging was found to be a preferable choice over L'-band in the inner, contrast-limited regime, whereas the opposite was found to be true in the outer, background-limited range. This conclusion is based on theoretical models that ultimately need to be confirmed through observations of known planets. The overlap occurs at a radius of 4$''$ in our images, a number that will depend on target brightness and integration time. Two PSF subtraction techniques were employed: regular active ADI as used previously, and a new combination of techniques, using PSF reference subtraction and aADI with the LOCI algorithm. While PSFR+aADI performs slightly better at large separations, the techniques are virtually indistinguishable for most of the contrast-limited regime. Using more than one PSF reference star may change this picture. In addition, the method of combining 4$\mu$m imaging and LOCI is also well suited for passive ADI, which has the potential to substantially enhance the performance even further.

In spite of the high sensitivities achieved in our images, we did not detect any potential companion candidate. Unless the known radial velocity companion to $\epsilon$ Ind A is a neutron star or even more exotic stellar remnant, the non-detection in all images implies that the system is probably older than 1 Gyr, possibly consistent with preliminary results presented by Cardoso et al. (2008). Furthermore, we can constrain the planet or brown dwarf mass to within approximately 5-20 $M_{\rm jup}$, the semi-major axis to $\sim$10-20 AU, and the inclination to $>$20$^{\rm o}$. An analysis based on astrometry from FK5 and \textit{Hipparcos} is consistent with such a companion. Given the high significance of the RV trend, the fact that we can exclude all stellar, white dwarf and high-mass brown dwarf companions, and the fact that exotic stellar remnants are rare, it seems very plausible that $\epsilon$ Ind A is one of the nearest stars to host a massive giant planet or very low-mass object. Furthermore, it is likely that this companion would be detectable through further imaging with either the presently available facilities, or facilities that come online in the relatively near future. Hence, $\epsilon$ Ind is a high-profile target for the study of substellar objects, even aside from the fact that it hosts the nearest binary brown dwarf.

Finally, we note that no sophisticated coronagraph adapted for observations beyond 3$\mu$m presently exists on any of the 8m-class or larger AO-assisted telescopes (although simple coronagraphs do exist, e.g. a Lyot coronagraph for NACO). The potential coronagraphic performance is intimately connected to the adaptive optics performance, which leads to an interest in coronagraphs in the context of 'extreme AO' facilities currently in development (e.g. Petit et al. 2008). However, given the fact that a demonstrated Strehl ratio in the range of 85\% can be reached even with NACO at 4$\mu$m, an 'extreme AO'-type performance in this particular wavelength range is available already today. The development of a coronagraph for this wavelength range could therefore be another promising avenue to further increase the near-future capacity of detecting extrasolar planets through direct imaging.

\section*{Acknowledgments}

The authors wish to thank Marten van Kerkwijk and Yanqin Wu for useful discussion. The study made use of the CDS and SAO/NASA ADS online services. M.J. is supported through the Reinhardt postdoctoral fellowship from the University of Toronto.

\label{lastpage}

\end{document}